\documentclass[a4paper]{PoS}
\usepackage[utf8]{inputenc}
\usepackage{amsmath}
\usepackage{longtable}
\usepackage{booktabs}
\usepackage{caption}
\usepackage{subcaption}

\title{A general data quality inspection for Gamma-Ray Bursts searches with HAWC}

\ShortTitle{DQ Inspection for GRBs searches with HAWC}

\author{\speaker{Cederik de León}, Humberto Salazar, Luis Villaseñor\\
        Benemérita Universidad Autónoma de Puebla.\\
        Av. San Claudio y 18 sur, colonia San Manuel Puebla México.\\
       E-mail: \email{cederik@gmail.com}}

\author{for the HAWC Collaboration\thanks{For a complete author list, 
see http://www.hawc-observatory.org/collaboration/icrc2017.php}}

\abstract{The High Altitude Water Cherenkov (HAWC) is a wide field-of-view 
gamma-ray observatory sensitive to gamma-rays in the 
300 GeV -- 100 TeV energy range, located in Mexico at an altitude of 4,100 m 
above sea level. The detector consists of 300 Water 
Cherenkov Detectors with a volume of 200,000 l each, having a footprint of 
22,000 $m^2$, a duty cycle of >95\% and a field of 
view (FoV) of 2sr \cite{Abeysekara2012641}. In this work we present a general 
data quality inspection of HAWC data for the years between 2014 and 2017 and 
also for 
two specific data periods selected on the base that GRB 150416A and GRB 160301A 
detected by FERMI occurred within the HAWC FoV, allowing to search for them 
in the HAWC data.}

\FullConference{35th International Cosmic Ray Conference – ICRC217-\\
		10--20 July, 2017\\
		Bexco, Busan, Korea}
	
\begin{document}

\section{Introduction}


Many events can produce excesses in the counts in the HAWC detector, which could have different
origins, such as: electrical atmospheric activity (i. e. thunder storms, lightning), noise due to the
experiment electronics and/or technical operations (i. e. maintenance and/or upgrades), unknown
fluctuations and signals from astrophysical objects (i. e. transients like 
GRBs) \cite{Abeysekara2012641}, \cite{0004-637X-800-2-78}.

The understanding of the data quality behavior of a certain period of interest, 
regardless the nature of the analysis to be performed, is essential. Lack of 
data or bad data periods will lead to a biased analysis or even make it 
impossible. A GRB analysis (like in \cite{0004-637X-800-2-78}) 
and other astrophysical phenomena (like in \cite{0004-637X-841-2-100}), 
needs data of enough quality to ensure unbiased 
variables to perform the search of signal associated to this astrophysical 
events.

For this work, we used the HAWC Quality and Monitoring Data 
Base (QMDB) and custom made computer scripts and programs in order to retrieve 
and analyze the related data quality variables.
Here we show a method to estimate the stability for the duration, the 
number of events and zenithal angle  from a subset of HAWC data (sub run) 
acquired and stored in a HAWC custom file format and electric field activity 
for selected data periods: 2014 -- 2017 and two specific dates 2015-04-16 and 
2016-03-01, where two GRBs occurred, triggered an external alert, were within 
the HAWC FoV and HAWC was in normal operation.

\section{HAWC datasets}

The data acquired by the HAWC DAQ \cite{Abeysekara2012641} is processed in a 
certain way that events are reconstructed and stored on--line, other 
information relevant for experiment, like scalers and TDC are as well stored. 
The data is organized in general blocks called runs, 
each run is labeled with a unique consecutive serial number that serves as 
identifier. A run contain sub runs (data files) as well identified by a name 
and a unique serial number. Sub run dataset file names are: 
reconstructed events (\textit{reco}), TDC data (\textit{tdc}), 
triggered data (\textit{trig}) and raw data (\textit{raw}), all of 
those files are generated by their own on--line process.

A typical run is expected to be a complete cycle of 24 hours with continuous 
data, nevertheless, due experiment operations (i. e. calibration, maintenance, 
repairs or outages) or external environmental condition (Weather 
conditions in general, particularly thunder storms or extreme electric field 
variations near to the HAWC Site) \cite{ABEYSEKARA2015125}, some runs last less 
than expected. When a run ends the experiment control start a new one (just 
in case of explicit operations of the experiment require to stop the data 
acquisition the automatic run stop--start process are not performed).

Each sub run file stores $\sim 125$ seconds of information; this 
is $\sim{691}$ sub run files per dataset ($\sim{3000}$ files per run, 
$>2\text{TB}$ per day).

\subsection{Quality Monitoring Data Base}

A way to track the experiment acquired data is implemented in a process that 
the statistics extraction, from the reconstructed files, is performed by a 
computer program \textit{diagnostics-rec-stats} that uses the HAWC AERIE 
Framework.

\textit{diagnostics-rec-stats} gets information related to 54 variables for 
a given reconstructed data file; some values are directly extracted 
from the file but in the majority are statistics computed from the entire 
data (events) contained into this file. The results are stored into the Quality 
Monitor Data Base (QMDB), a MySQL data base. The data can be accessed for the 
HAWC collaboration.

There is two main groups of data sets with exactly the same 
variable definition in QMDB, the HAWC On--line and the 
HAWC Production. The main distinction between this datasets is the way that the 
HAWC data are reconstructed.

\section{GRB external alerts and HAWC Data}

HAWC have a data base that stores information related to alerts triggered from 
other experiments (external alerts). We retrieve information from external 
alerts in order to explore the possibility of analysis. The selection was done 
using the conditions shown in Table \ref{tab:GRBDataconditions}.

\begin{table}[th!]
 \begin{center}
  \begin{tabular}{*2c}
 \toprule
  \multicolumn{2}{c}{Conditions} \\
 \midrule
   Date range & 2014--2017 (May) \\
   Zenith angle (HAWC) & $[ 0\le\theta\le 45 ]$ degrees\\
   Instrument & FERMI, SWIFT\\
 \bottomrule
  \end{tabular}
 \end{center}
\caption[GRBs external alerts conditions]{GRBs external alerts 
conditions applied to search and retrieve its relevant information 
as: Instrument, Date and Time (UTC), Zenith Angle (relative to HAWC)}
\label{tab:GRBDataconditions}
  \vspace{-1.5em}
\end{table}
 
\begin{figure}[th!]
 \begin{center}
  \includegraphics[scale=.35]{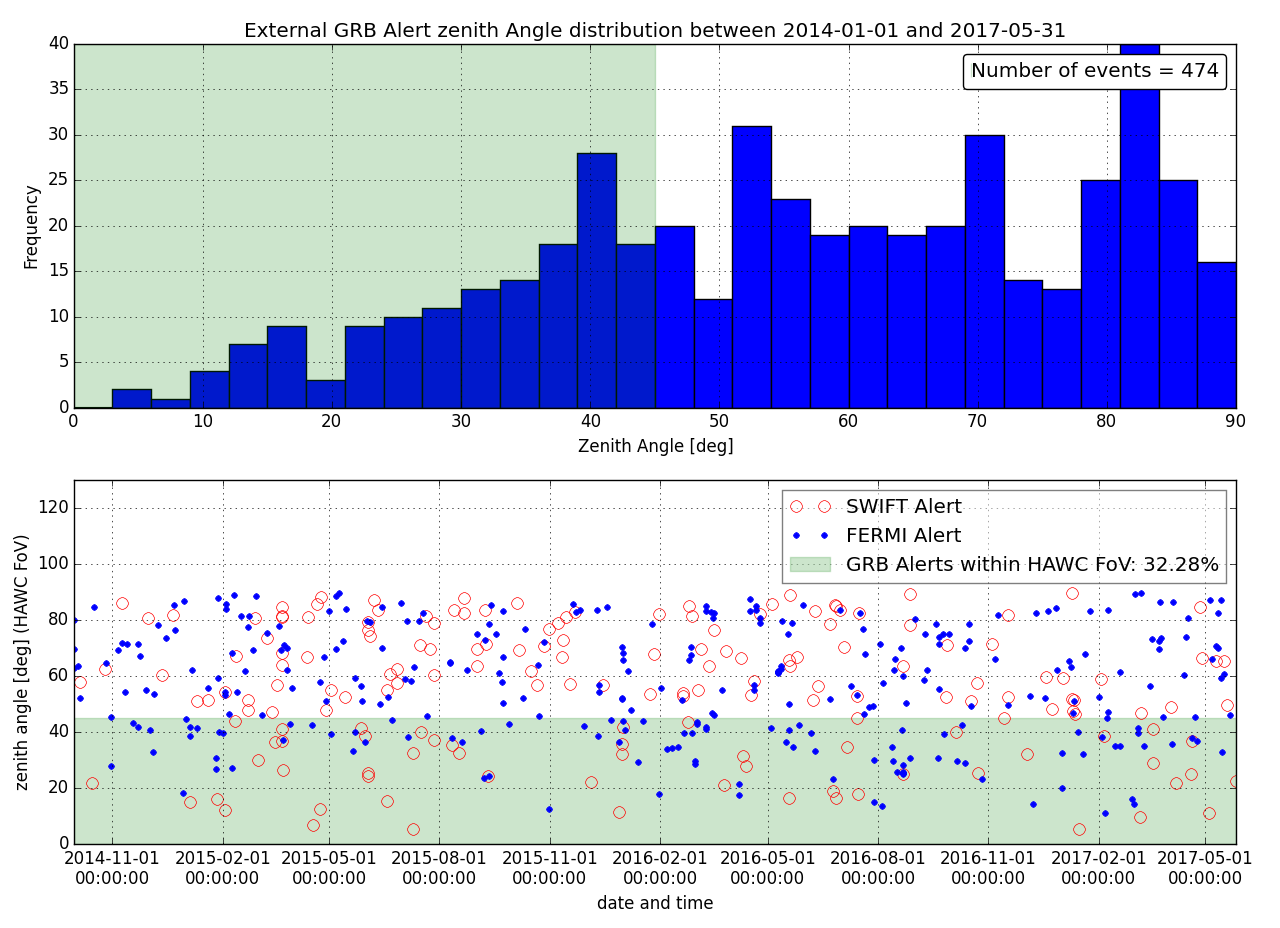}
  \end{center}
 \caption[GRB external alerts distribution]{Distribution of the GRB 
external alerts within $0\le\theta\le90$ degrees relative to the HAWC zenith 
(Top plot). GRB external alerts, the open red circles represents the SWIFT 
alerts, the blue dots represents the FERMI alerts, the HAWC instantaneous Field 
of View (FoV) is in the $0\le\theta\le45$ degrees in shaded region (Bottom 
plot).}
\label{fig-grbalertdist2014-2017}
  \vspace{-.5em}
\end{figure}

After applying the conditions mentioned above, we got a the fraction of 
external alerts within the HAWC FoV as shows in 
Table \ref{table:GRBFraction}, the plot in Figure 
\ref{fig-grbalertdist2014-2017} shows the GRB external alerts distribution 
within $0\le\theta\le90$ degrees, relative to HAWC (top plot). The open red 
circles represents the 
SWIFT alerts, blue dots represents the FERMI alerts, the HAWC instantaneous 
Field of View (FoV) is in the $0\le\theta\le45$ degrees in shaded region (bottom 
plot).
 
\begin{table}[th!]
\begin{center}
\begin{tabular}{|c|c|}
\hline
 Year & Fraction of GRB within HAWC FoV\\
\hline
\hline
 2014 & $\sim 0.14$\\
 \hline
 2015 & $\sim 0.26$\\
\hline
 2016 & $\sim 0.23$\\
\hline
 2017 & $\sim 0.39$\\
 \hline
\end{tabular}
\end{center}
\caption[HAWC GRB external alerts by year]{Fraction of GRB external alerts 
within the FoV ($\#\text{external alerts (Within HAWC
FoV) $0\le\theta\le45\text{ [degrees]}$}/{\#\text{external alerts } 
0\le\theta\le90\text{ [degrees]}}$)}
\label{table:GRBFraction}
\end{table}

For each GRB external alert between the date range (see Table 
\ref{tab:GRBDataconditions}) we ask for a data duration of 24 hours, with a 
tolerance $\le 0.03$, will be used for the GRB external alert and always will 
contain the triggered alert Date--Time. In this work we centered the GRB 
external alert, starting the dataset period 12 hours before the GRB external 
alert trigger and 12 after the GRB external alert trigger. 

\begin{table}[h!]
 \begin{center}
  \begin{tabular}{*2c}
 \toprule
  \multicolumn{2}{c}{Cuts} \\
 \midrule
Duration required & $\pm12$ hours (GRB alert centered)\\
Duration ratio ($\frac{\text{data duration}}{\text{duration 
required}}$) & $\ge 0.97$ \\
Data duration stability ($\sigma / \mu$ ratio) & $< 0.05$ \\
Number of events stability ($\sigma / \mu$ ratio) & $< 0.05$ \\
Zenith angle stability & $20<\text{med}(\theta)<25$\\
 \bottomrule
  \end{tabular}
 \end{center}
\caption[GRBs external alerts dataset quality cuts]{GRBs external alerts 
dataset quality cuts applied in the period related at every external alert 
between 2014 and 2017 (may) }
\label{tab:GRBDataCuts}
\end{table}

For the sub run duration and number of events stability we used the  
$\sigma/\mu$ ratio, this value was estimated using a computer simulation 
considering that the data follows a uniform distribution, without 
discontinuities. We made 1000 simulations for a 24 hours of data. Every 
simulations contains a set of $\sim{691}$ sub run with their duration values  
$125\pm0.001$ in seconds, noise was introduced randomly and consist a fraction 
of the total number of sub run entries: $[0.0025-0.05]$.  The ``number of 
events'' in a sub run is considered uniform as well in the 24 hours period and 
its simulation was performed in the same way, described above. 

$\sigma/\mu$ ratio values $\ge{0.05}$ are 
considered as poor quality period data set, it implies a ``noisy'' dataset 
(noise is considered as extreme values far from the mean and in a proportion 
$N/S\ge0.0075$).

We seek for the duration ratio condition mets, if 
not, the requested total duration are then reduced symmetrically (e. g. 1000 
seconds both sides, before and after. This is an arbitrary value and could 
change), this process continues until the ratio ($\ge 0.97$) is reached. 
If the duration ratio is never reached the event is considered  
as ``poor data quality''.


\section{HAWC data quality inspection for GRBs external alerts}

\begin{figure}[ht!]
 \begin{subfigure}{.5\textwidth}
  \includegraphics[scale=.2]{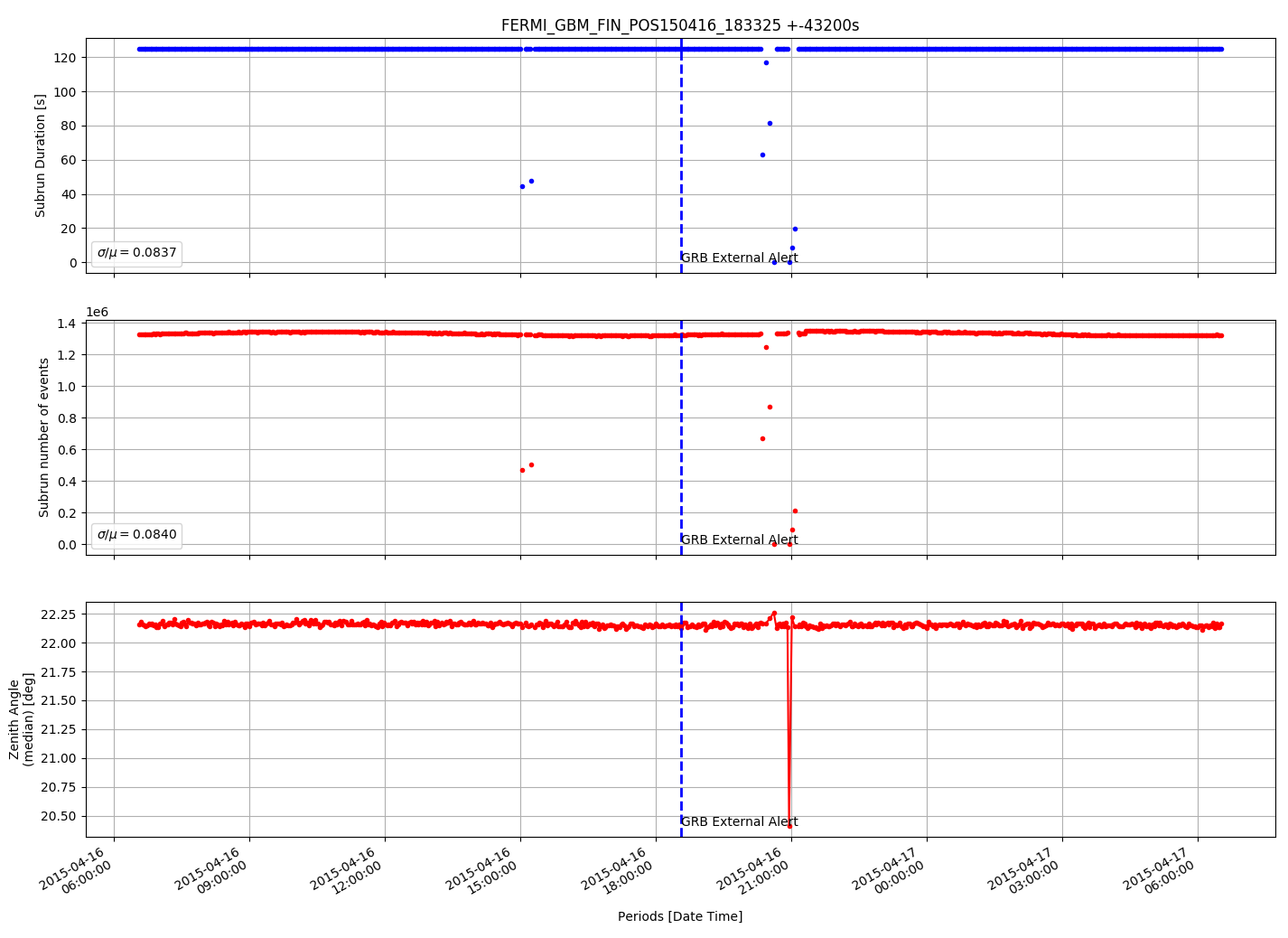}
  \caption[GRB 150416A data quality inspection]{}
  \label{fig:GRB150416ADQ}
 \end{subfigure}
 \begin{subfigure}{.5\textwidth}
   \includegraphics[scale=.2]{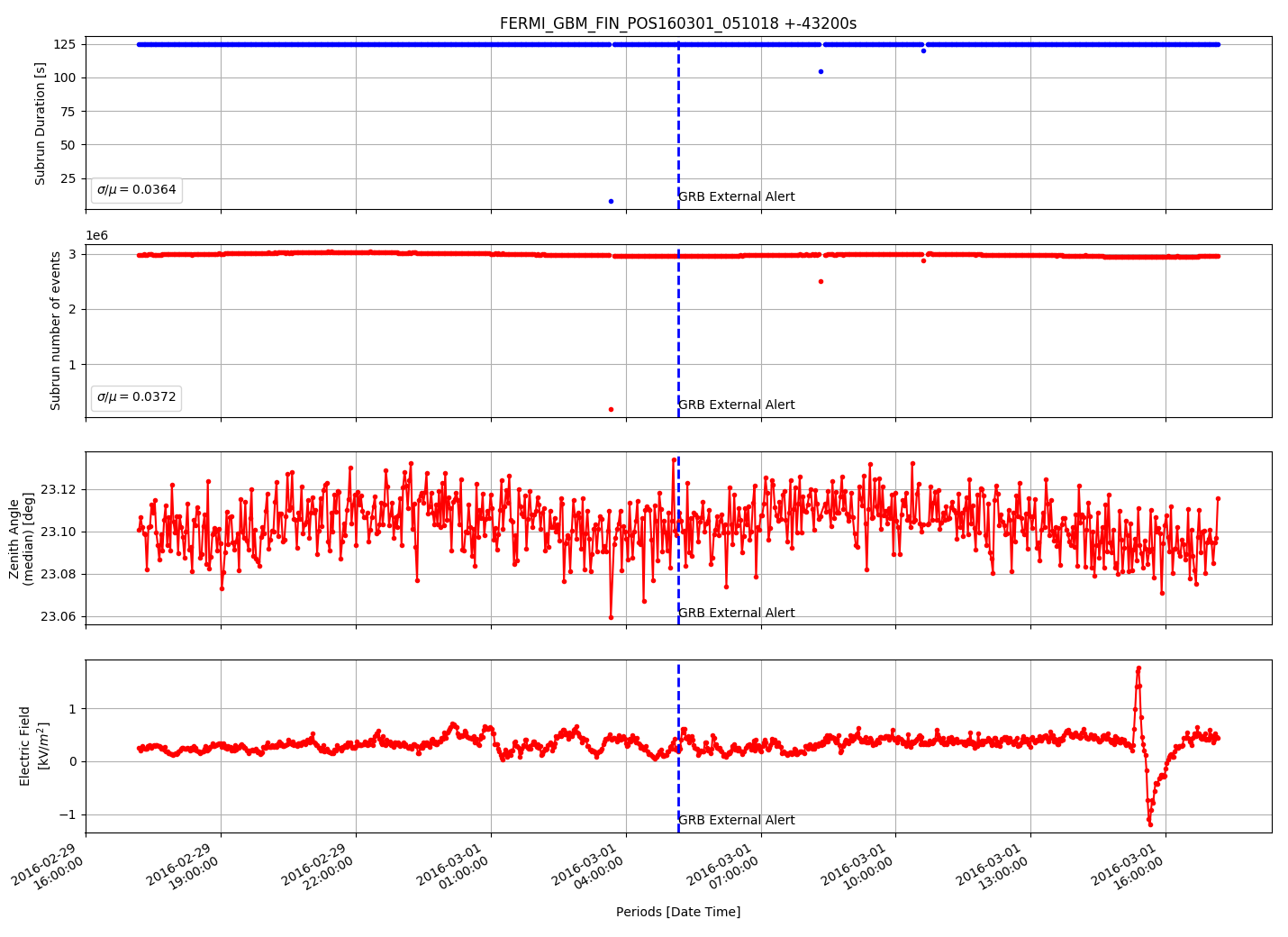}
   \caption[GRB 160301A data quality inspection]{}
   \label{fig:GRB160301ADQ}
 \end{subfigure}
\caption[GRB150416A and GRB160301A data quality inspection]{ 
(\ref{fig:GRB150416ADQ}) GRB150416A data quality inspection: (Top to down) sub 
run duration, number of events per sub run, sub runs median of the zenith 
angle, for this period HAWC has no electric field records. 
(\ref{fig:GRB160301ADQ}) GRB160301A DQ Inspection: (Top to down) sub run 
duration, number of events per sub run, sub runs median of the 
zenith angle, electric field activity.}
\label{fig:GRB150416A-160301ADQ}
\end{figure}

We selected and inspected the 24 hours data periods for two GRB external 
alerts: GRB150416A and GRB160301A, their results are shown in Table 
\ref{table:GRB150416A-160301A}, Figure \ref{fig:GRB150416A-160301ADQ} contains 
information for every sub run in the GRBs 24 hours data period.

\begin{table}[h!]
\begin{tabular}{|c|c|c|c|c|}
\hline
Instrument & date and time (UTC) & Best duration found [s] & 
duration ratio & $\sigma/\mu$--value \\
 \hline
 FERMI GBM & 2015-04-16 18:33:25 & 43200 & $\sim 0.97 $ & $\sim 0.08 $ \\
 FERMI GBM & 2016-03-01 05:10:18 & 43200 & $\sim 0.99 $ & $\sim 0.03$ \\
 \hline
\end{tabular}
\caption[GRB150416A and GRB160301A data quality inspection]{GRB150416A and
GRB160301A data quality inspection}
 \label{table:GRB150416A-160301A}
\end{table}

\begin{figure}[ht!]
 \begin{subfigure}{.5\textwidth}
  \includegraphics[scale=.2]{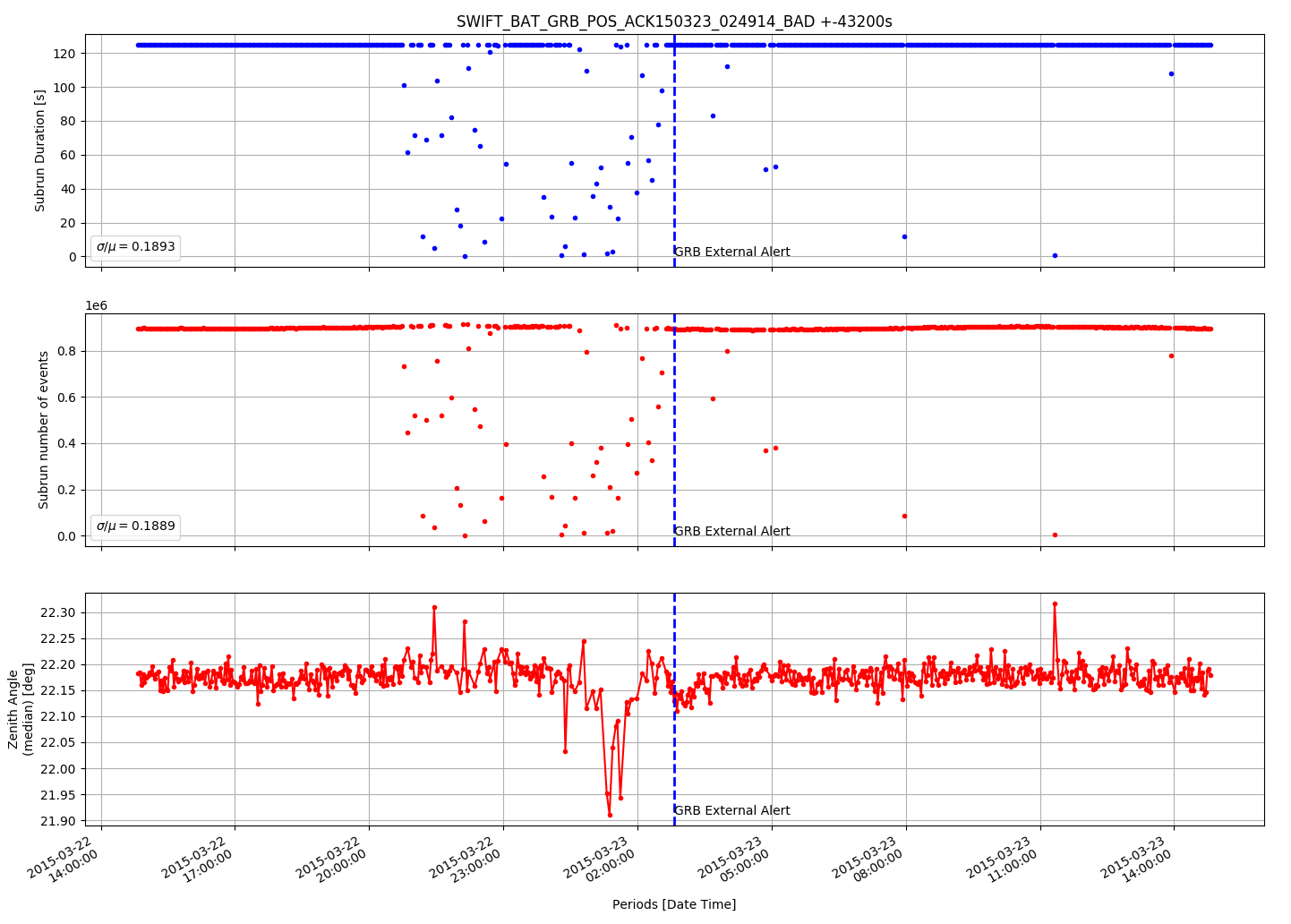}
   \caption[GRB 150323A data quality inspection]{}
   \label{fig:GRB150323ADQ}
  \end{subfigure}
 \begin{subfigure}{.5\textwidth}
   \includegraphics[scale=.2]{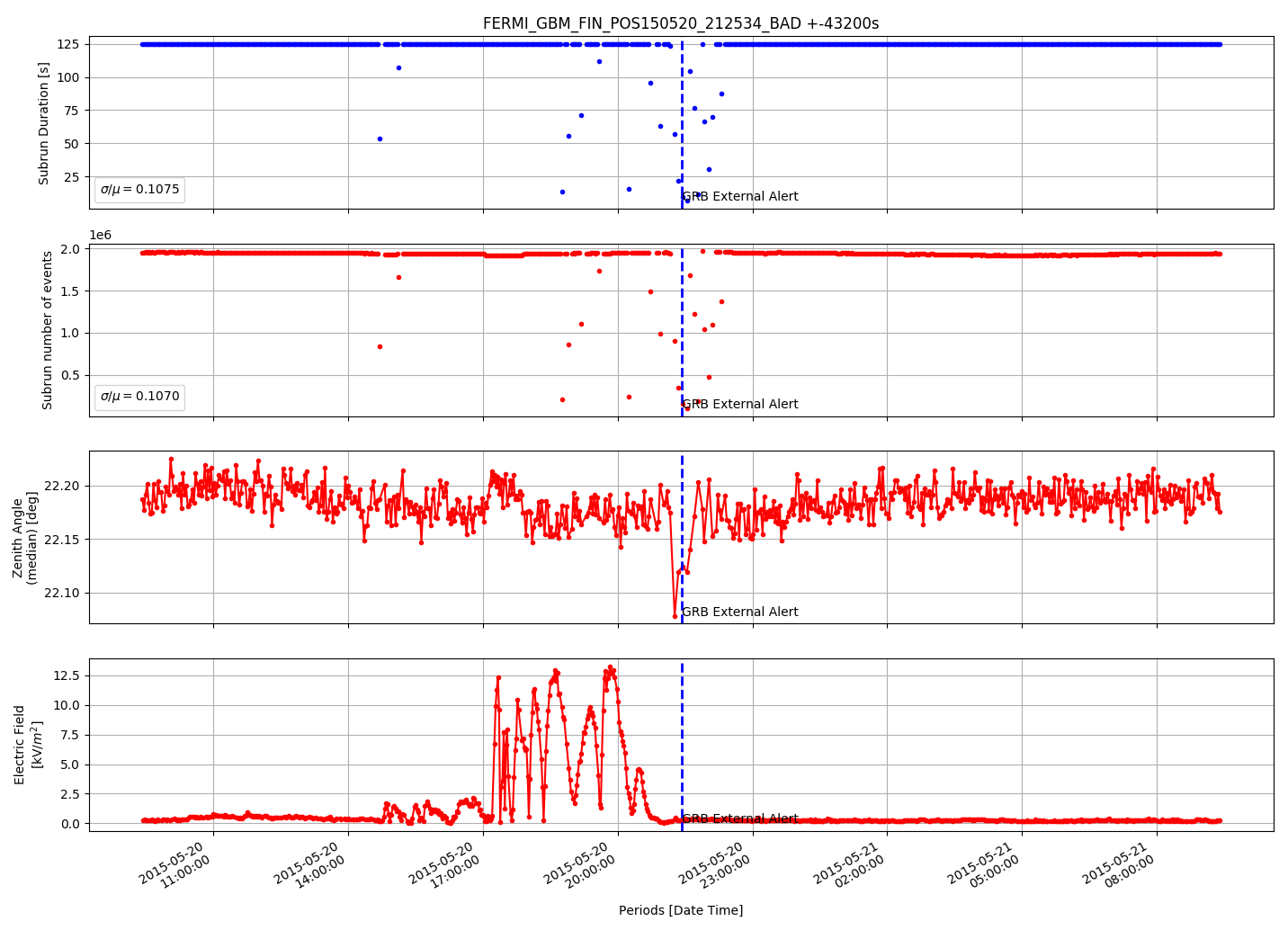}
   \caption[GRB150520A data quality inspection]{}
   \label{fig:GRB150520ADQ}
 \end{subfigure}
 \caption[GRB150323A and GRB150520A DQ inspection ]{(\ref{fig:GRB150323ADQ}) 
GRB150323A data quality inspection: (Top to down) sub run duration, number of 
events per sub run, sub runs median of the zenith angle, for this period HAWC 
has no electric field records. (\ref{fig:GRB150520ADQ}) GRB150520A data quality 
inspection. Top to down: sub run  duration, number of events per sub 
run, sub runs median of the zenith angle,  electric field activity.}
\label{fig:GRB150323A-150520ADQ}
\end{figure}

Comparing datasets, we seek for another two (of nine) ``poor data 
quality'' GRBs external alerts: GRB150323A and GRB150520A, their results are 
shown in Table \ref{table:GRB150323A-150520A}, Figures 
\ref{fig:GRB150323A-150520ADQ} 
contains information for every sub run in the GRBs 24 hours data period.

\begin{table}[th!]
\begin{tabular}{|c|c|c|c|c|}
\hline
Instrument & date and time (UTC) & Best duration found [s] & 
duration ratio & $\sigma/\mu$--value \\
 \hline
 SWIFT BAT & 2015-03-23 02:49:14 & ND & ND & $> 0.18 $ \\
 FERMI GBM & 2015-05-20 21:25:34 & ND & ND & $> 0.10 $ \\
 \hline
\end{tabular}
\caption[GRB150323A and GRB150520A data quality inspection]{GRB150323A and 
GRB150520A data quality inspection (ND = No--Data).}
 \label{table:GRB150323A-150520A}
\end{table}

The environmental conditions at the HAWC site have special importace, every 
GRB external alert and its associate data period includes, if available, the 
electric field (EF) related data, in order to consider and explore a possible 
bias on data acquisition; we observed that only on 
periods of extreme electric field variations (i. e. thunder storms near to the 
site) the data stability is affected.


Figures \ref{fig:GRB150323ADQ} and \ref{fig:GRB150520ADQ} shows
the sub run duration and number of events are spread and the median of zenith 
angle drops dramatically, for the GRB150520A an increment in 
the electric field is present.

\section{GRB150416A and GRB160301A analysis trial with data quality 
inspection}

As a preliminary analysis we applied a Moving Average (MA) analysis on a set 
of TDC files belonging to GRB external alerts for GRB150416A and GRB160301A; 
the data duration stability behavior for the TDC files shows the same 
behavior as mentioned in Table \ref{table:GRB150416A-160301A}. We use the same 
data period of 24 hours to perform a preliminary Moving Average analysis on 
them. We also used the information from Table \ref{table:GRB150323A-150520A} 
(poor datasets) to perform the same analysis, nevertheless due their poor data 
quality (mainly in duration and stability) we did not get any results.

\begin{figure}[ht!]
 \begin{subfigure}{.5\textwidth}
  \includegraphics[scale=.16]{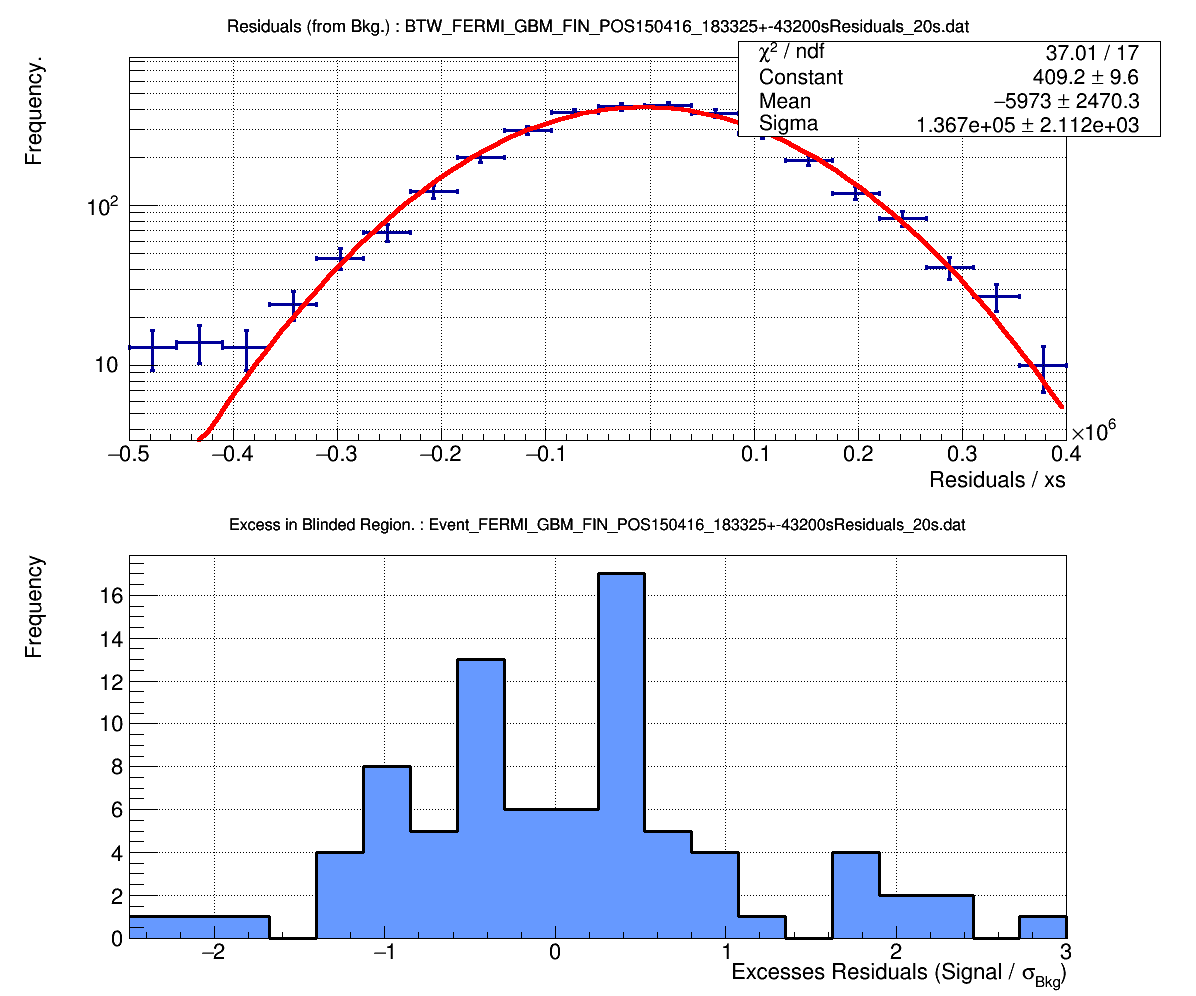}
   \caption[GRB150416A MA Background estimation]{}
   \label{fig:GRB150416AMABkg}
  \end{subfigure}
 \begin{subfigure}{.5\textwidth}
  \includegraphics[scale=.25]{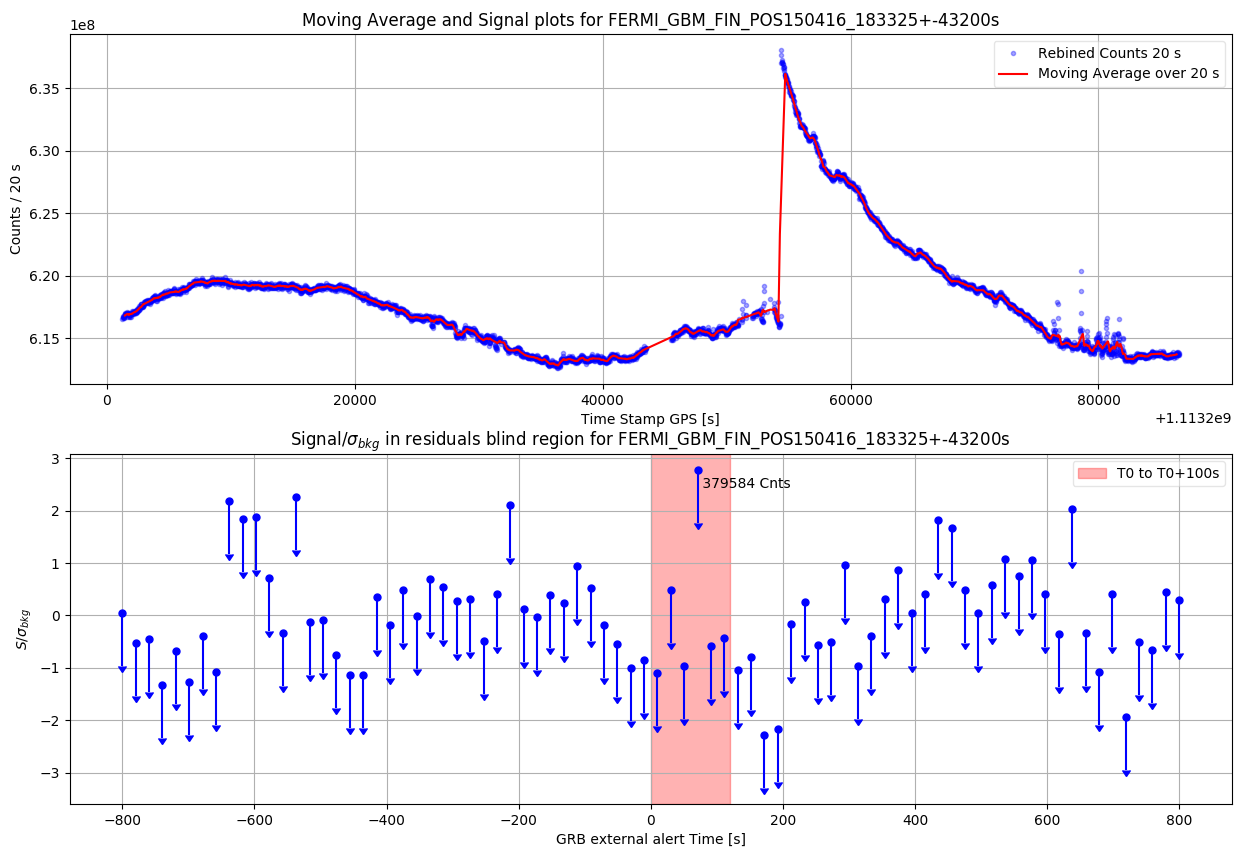}
   \caption[GRB150416A Moving Average signaal inspection]{}
   \label{fig:GRB150416AMASignal}
  \end{subfigure}
\caption[GRB150416A Background and MA]{GRB150416A: (\ref{fig:GRB150416AMABkg}) 
Background estimation, gaussian fit (top), excesses histogram (bottom), 
(\ref{fig:GRB150416AMASignal}) Moving Average signal (top), excesses 
distribution, shaded area are $T_{0}+100$ (bottom),} 
\label{fig:GRB150416AMARA}
\end{figure}

For the GRB150416A, Figure \ref{fig:GRB150416AMARA} shows a preliminary 
analysis using the MA (this trial version uses a rebin of $x=20s$ 
and considered 10 samples for the Moving Average calculation, Figure 
\ref{fig:GRB150416AMABkg} background estimation (top), excesses histogram in 
the blind region $\pm1000$s (bottom). Figure 
\ref{fig:GRB150416AMASignal} signal behavior in the dataset used to estimate 
the background (top), excesses distribution in 
the blind region $\pm1000$s, shaded area are $T_{0}+100$ (bottom). In this 
signal we saw an artifact near the second $55000 + 1.1132\times10^{9}$, this need to be 
studied as an special case ($T_{0}$ = Date and time when the GRB triggered the external alert reported).

\begin{figure}[ht!]
 \begin{subfigure}{.5\textwidth}
  \includegraphics[scale=.16]{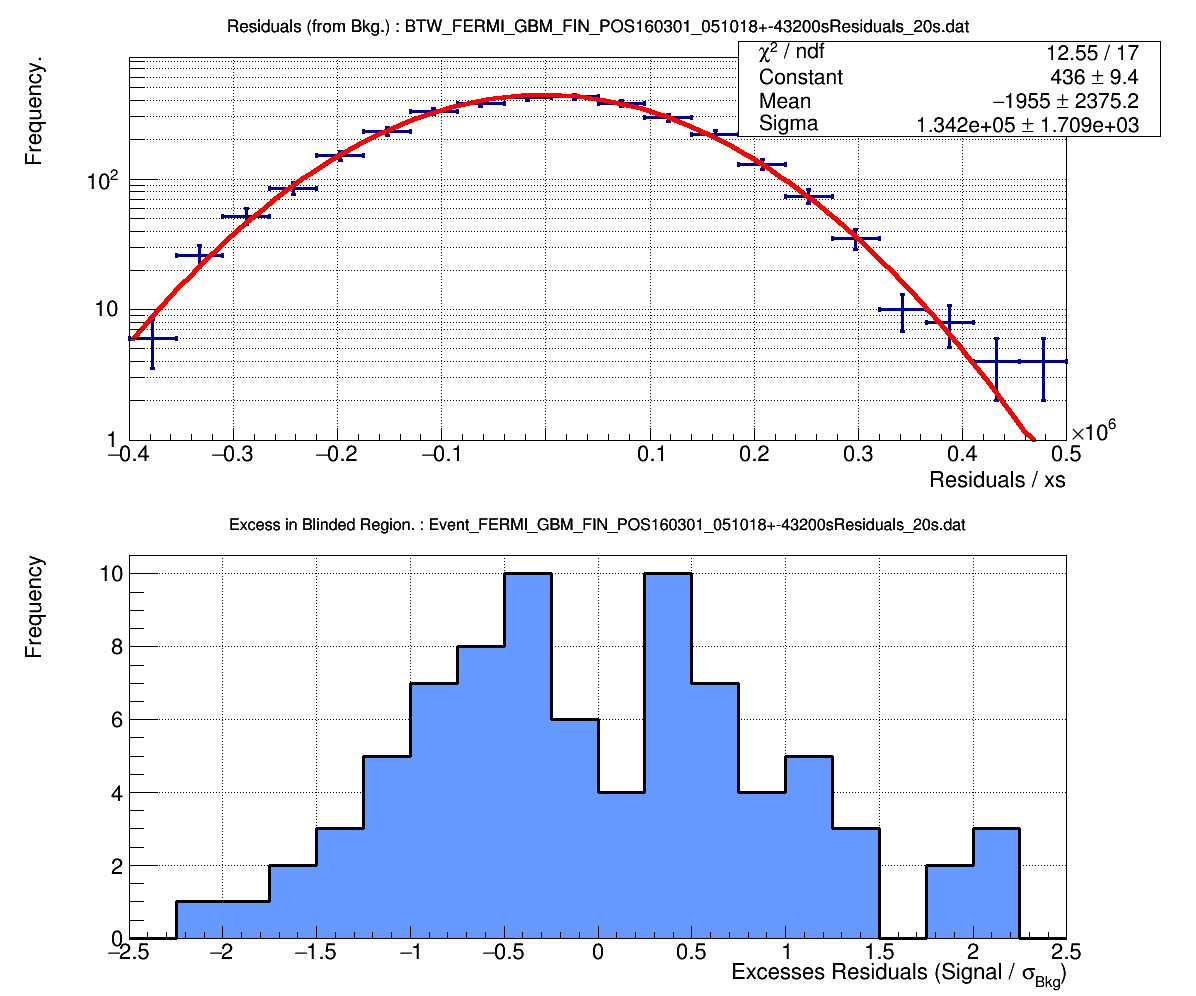}
   \caption[GRB160301A MA Background estimation]{}
   \label{fig:GRB160301AMABkg}
  \end{subfigure}
 \begin{subfigure}{.5\textwidth}
  \includegraphics[scale=.25]{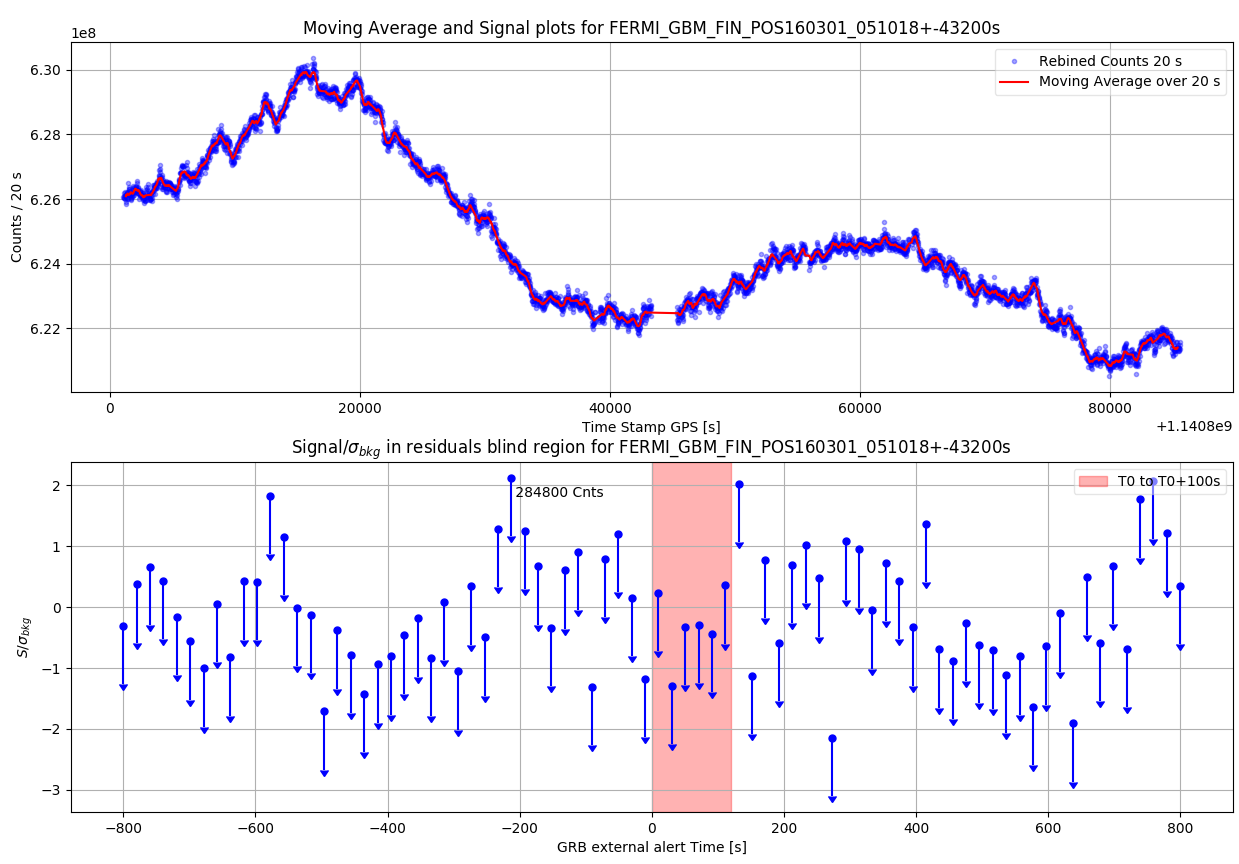}
  \caption[GRB160301A Moving Average signal inspection]{}
   \label{fig:GRB160301AMASignal}
 \end{subfigure}
\caption[GRB160301A Background and MA]{GRB160301A: (\ref{fig:GRB160301AMABkg}) 
Background estimation, gaussian fit (top), excesses histogram (bottom), 
(\ref{fig:GRB160301AMASignal}) Moving Average Moving Average signal (top), 
excesses distribution, shaded area are $T_{0}+100$ (bottom)}
\label{fig:GRB160301AMARA}
\end{figure}

For the GRB160301A, Figure \ref{fig:GRB160301AMARA} shows a preliminary 
analysis using the MA (this trial uses a rebin of $x=20s$ 
and considered 10 samples for the Moving Average calculation, Figure 
\ref{fig:GRB160301AMABkg} background estimation (top), excesses histogram in 
the blind region $\pm1000$s (bottom). Figure \ref{fig:GRB160301AMASignal} the signal behavior in the dataset used to 
estimate the background (top), excesses distribution in 
the blind region $\pm1000$s, shaded area are $T_{0}+100$ (bottom) ($T_{0}$ = 
Date and time when the GRB triggered the external alert, reported by its 
experiment).


%
\section{Summary}

The HAWC experiment datasets analyzed for the GRBs external alerts between 
2014--2017 (may) has been explored and shows a reliable source of data to 
perform analysis with them.

Between 2014--2017 (may) a total of 147 GRB External alerts in the HAWC FoV has 
been registered, 26 of them ($<18\%$) were marked with this method and its cuts 
 as ``poor data quality'' due a lack of duration, 9  ($<7\%$) never reached the requested duration ratio nor the 
$\sigma/\mu$ threshold even with the period reduction process applied.

This general data quality inspection method, for the HAWC experiment, is an 
easy and fast way to find stable data periods. Environmental conditions needs to be taken into account, if activity of 
this kind is present in the selected periods, it should be treated in the most 
appropriate way for their purposes.

\section{Acknowledgments}

We acknowledge the support from: the US National Science Foundation (NSF); the
US Department of Energy Office of High-Energy Physics; the Laboratory Directed
Research and Development (LDRD) program of Los Alamos National Laboratory;
Consejo Nacional de Ciencia y Tecnolog\'{\i}a (CONACyT), M{\'e}xico (grants
271051, 232656, 260378, 179588, 239762, 254964, 271737, 258865, 243290,
132197), Laboratorio Nacional HAWC de rayos gamma; L'OREAL Fellowship for
Women in Science 2014; Red HAWC, M{\'e}xico; DGAPA-UNAM (grants IG100317,
IN111315, IN111716-3, IA102715, 109916, IA102917); VIEP-BUAP; PIFI 2012, 2013,
PROFOCIE 2014, 2015;the University of Wisconsin Alumni Research Foundation;
the Institute of Geophysics, Planetary Physics, and Signatures at Los Alamos
National Laboratory; Polish Science Centre grant DEC-2014/13/B/ST9/945;
Coordinaci{\'o}n de la Investigaci{\'o}n Cient\'{\i}fica de la Universidad
Michoacana. Thanks to Luciano D\'{\i}az and Eduardo Murrieta for technical 
support; National Laboratory; Polish Science Centre grant 
DEC-2014/13/B/ST9/945; The authors thankfully acknowledge the computer 
resources, technical expertise and support provided by the Laboratorio Nacional 
de Supercómputo del Sureste de México, CONACYT Network of national laboratories.
Special thanks to: Eduardo Moreno and Humberto Martínez.

\end{document}